\documentclass[sigconf]{acmart}

\AtBeginDocument{%
  }

\setcopyright{acmlicensed}
\copyrightyear{2026}
\acmYear{2026}
\acmDOI{XXXXXXX.XXXXXXX}
\acmConference[Conference acronym 'XX]{XXX}{June 03--05,
  2018}{Woodstock, NY}
\acmISBN{978-1-4503-XXXX-X/2026/10}

\acmConference[Conference acronym 'XX]{XXX}{June 03--05,
  2018}{Woodstock, NY}

\usepackage{amsmath}

\usepackage{amssymb}
\usepackage{booktabs}
\usepackage{multirow}
\usepackage{xcolor}
\usepackage{subcaption}
\usepackage{tikz}
\usetikzlibrary{arrows.meta,positioning,fit,calc,decorations.pathreplacing}

\newcommand{\ours}{\textsc{PrefixMem}}

\begin{document}

\title{LLMs Need Encoders for Semantic IDs Too}

\author{Xiangyi Chen}
\email{xiangyichen@pinterest.com}
\affiliation{%
  \institution{Pinterest}
  \country{United States}}

\author{Zelun Wang}
\email{zelunwang@pinterest.com}
\affiliation{%
  \institution{Pinterest}
  \country{United States}}

\author{Xinyi Li}
\email{xinyi@pinterest.com}
\affiliation{%
  \institution{Pinterest}
  \country{United States}}

\author{Yi-Ping Hsu}
\email{yhsu@pinterest.com}
\affiliation{%
  \institution{Pinterest}
  \country{United States}}

\author{Jaewon Yang}
\email{jaewonyang@pinterest.com}
\affiliation{%
  \institution{Pinterest}
  \country{United States}}

\author{Jiajing Xu}
\email{jiajing@pinterest.com}
\affiliation{%
  \institution{Pinterest}
  \country{United States}}

\begin{abstract}
Multimodal LLMs use dedicated encoders to bridge non-language modalities
(vision encoders for images, depth models for audio codec tokens) because
raw token embeddings alone cannot capture modality-specific structure.
We argue that Semantic IDs (SIDs), the hierarchical codes used in generative
recommendation, constitute another such modality: a SID level token's meaning
depends on its prefix context, yet current systems simply add SID tokens to
the vocabulary and rely on training to learn these context-dependent meanings
from scratch.

We propose \ours{}, a lightweight SID encoder based on prefix $n$-gram memory
tables that provides the LLM with structured, prefix-conditioned
representations at SID token positions. Like vision encoders in multimodal
LLMs, \ours{} can be pre-trained independently and then attached to any LLM
for joint training.
We evaluate on large-scale data from Pinterest across multiple LLM families
and show that \ours{} improves deepest-level SID accuracy by up to 46\% relative
and full-SID retrieval recall by up to 22\% relative at matched training compute.
The encoder's benefit concentrates on hard examples where greedy decoding
fails, with up to 77\% relative accuracy gains, confirming that SID tokens
benefit from a dedicated encoder just as other non-language modalities do.
\end{abstract}



\maketitle

\section{Introduction}
\label{sec:intro}
Generative recommendation produces item identifiers token by token from a
single
model~\cite{tay2022transformer,rajput2023recommender,zhai2024hstu,he2026plum,deng2025onerec}.
Items are represented as Semantic IDs (SIDs), hierarchical codes from
RQ-VAE~\cite{lee2022autoregressive} that enable knowledge transfer across
semantically similar
items~\cite{singh2024better,letter2024,cost2024,eager2024,etegrec2025,snap_sid2026,pinrec2025,mtgr2025}. Our goal is an LLM small enough to serve production traffic that understands
the SID hierarchy for a billion-scale item corpus.

\begin{figure}[t]
  \centering
  \includegraphics[width=\linewidth]{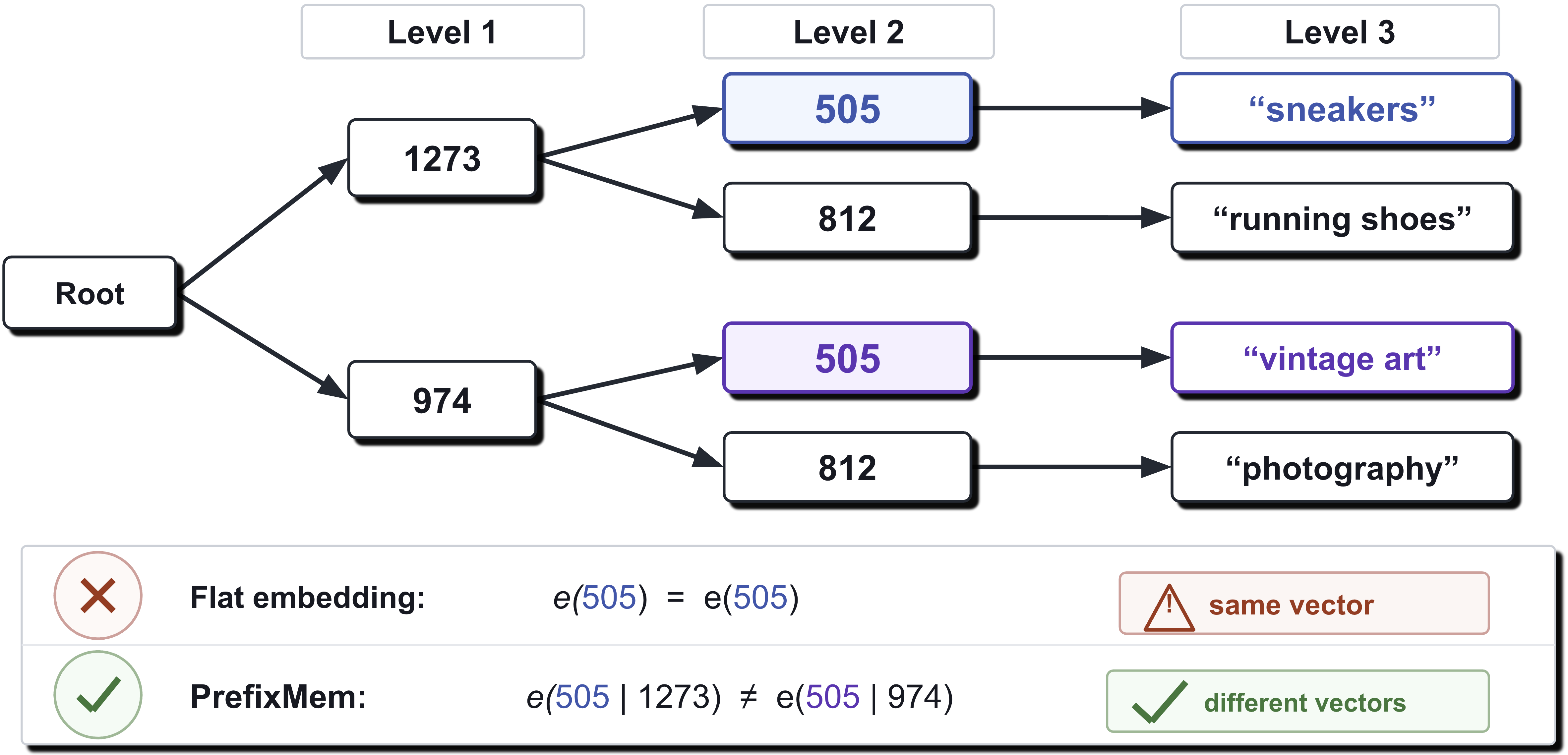}
  \caption{The same SID code means different things under different prefixes. \ours{} provides prefix-conditioned vectors; flat embeddings cannot distinguish them.}
  \label{fig:motivation}
\end{figure}

SIDs are a hierarchical vocabulary entirely foreign to a pre-trained LLM.
A code's meaning depends on its prefix: code 505 after prefix 1273 may index
sneakers, while the same 505 after prefix 974 indexes vintage art (Fig.~\ref{fig:motivation}). At
level~$\ell$ there are $K^{\ell-1}$ possible prefix contexts, making the
learning problem combinatorially large and sparse. Intermediate levels further
suffer from the hourglass phenomenon~\cite{kuai2024hourglass}, where code
distributions collapse at middle hierarchy levels. At billion-scale, every item
needs sufficient training exposure for its codes to be learned in context,
making the problem expensive even before considering multiple passes.

Existing methods treat SID codes as flat vocabulary tokens: the same code
receives the same embedding regardless of prefix
context~\cite{zhou2025openonerec,zheng2024lcrec,sidreasoner2026,liu2025onerec}.
Smart initialization~\cite{chen2026gti,minixhofer2022wechsel} provides a better
starting point for new embeddings but does not address the prefix-dependent
meaning problem and has only been demonstrated at small corpus scale. Constrained
decoding~\cite{su2026static,hou2025rpg} prevents hallucinated SID combinations
at inference time but does not improve the model's understanding of the
hierarchy. Massive continual pre-training~\cite{he2026plum} works but requires
hundreds of billions of tokens and couples SID knowledge to a specific LLM
checkpoint.

The multimodal LLM community solved analogous problems with dedicated encoders.
Vision encoders project images into the LLM's input
space~\cite{liu2023visual}. Audio LLMs face the same structural challenge:
speech is tokenized via RQ-VAE into hierarchical codec codes, structurally
identical to SIDs, and dedicated depth transformers handle the fine codebook
levels~\cite{defossez2024moshi}. We apply this principle to recommendation.

We propose \ours{}, a lightweight SID encoder based on hash-based prefix
$n$-gram memory tables~\cite{zheng2025enhancing,cheng2026conditional}. For each SID level, the encoder hashes the preceding
codes into a prefix-conditioned vector and adds it to the token embedding. The
LLM thus receives a different representation for the same code depending on its
prefix. The encoder is pre-trained on the item corpus independently, so the LLM can
focus on using the structure rather than discovering it from scratch. The encoder parameters
are sparse lookup tables, transferable across LLM families. The approach is
complementary to constrained decoding and initialization strategies, which can
be applied on top.

We evaluate \ours{} on billion-scale data from Pinterest across multiple LLM families and summarize our contributions:
\begin{itemize}
    \item We frame SIDs as a hierarchical vocabulary requiring a dedicated
    encoder, analogous to vision and audio encoders in multimodal LLMs.
    \item We propose \ours{}, a prefix $n$-gram memory module that provides
    prefix-conditioned representations at SID token positions. At matched
    compute, \ours{} improves deepest-level SID accuracy by up to 46\% relative
    and full-SID retrieval recall by up to 22\% relative. The encoder's benefit concentrates on ``hard'' examples where
    greedy decoding fails: on these cases, accuracy improves by up to 77\%
    relative.
    \item  The encoder also improves SID-to-text grounding (BLEU +33\%),
    confirming it helps the LLM understand what SIDs mean, not just predict
    the next code.
    \item Encoder pre-training is cheap (sparse table lookups, no transformer
    FLOPs) and decoupled from the LLM: the same encoder transfers across LLM
    families (Qwen, Llama, Gemma). A 0.6B model with \ours{} outperforms a
    4B model without it on SID prediction.
    \item We compare alternative encoder architectures (hash memory vs.\
    transformer) to identify what design properties matter for SID encoding.
\end{itemize}


\section{Related Work}
\label{sec:related}

\paragraph{Generative Recommendation with Semantic IDs.}
TIGER~\cite{rajput2023recommender} introduced encoding items as hierarchical
codes via RQ-VAE for generative retrieval. Subsequent work improved SID
construction through collaborative
signals~\cite{letter2024,cost2024}, behavioral
modeling~\cite{eager2024,etegrec2025}, and better
parameterization~\cite{singh2024better,zheng2025enhancing}, with industrial
deployment at YouTube~\cite{he2026plum},
Kuaishou~\cite{deng2025onerec,liu2025onerec,zhou2025onerec,zhou2025onerecv2,zhou2025openonerec},
Snapchat~\cite{snap_sid2026}, Meituan~\cite{mtgr2025}, and
Pinterest~\cite{pinrec2025}. On the modeling side,
P5~\cite{geng2022p5} first cast recommendation as language processing,
PLUM~\cite{he2026plum} scaled LLM-based generation to billions of items via
continual pre-training, and further work explored SID-language
alignment~\cite{zheng2024lcrec,sidreasoner2026,liang2026generative}. All these
systems add SID codes as flat vocabulary tokens; our encoder provides
prefix-conditioned representations that any of them could adopt.

\paragraph{Encoders and Memory Augmentation for LLMs.}
When LLMs are extended to new modalities, dedicated encoders bridge the
representation gap: vision encoders for
images~\cite{liu2023visual} and depth transformers for hierarchical audio codec
codes~\cite{defossez2024moshi}, which are structurally identical to SIDs
(hierarchical codes from RQ-VAE). Separately, hash-based memory has emerged as
a way to augment LLMs with sparse, high-capacity lookup: Zheng et
al.~\cite{zheng2025enhancing} used prefix $n$-gram hash tables to parameterize
SID embeddings in a DLRM setting, and Cheng et
al.~\cite{cheng2026conditional} proposed scalable conditional memory for
general LLMs. Our work combines both lines: we treat SIDs as a modality
requiring a dedicated encoder, implemented via hash-based prefix memory, and
evaluate it as a pre-trainable module across multiple LLM families.

\section{Method}
\label{sec:method}
\subsection{Training LLMs with Semantic IDs}
\label{sec:method:setup}

Following recent work~\cite{he2026plum,liu2025onerec}, we adapt a pre-trained
LLM to generate and understand Semantic IDs. Each item in the catalog is
encoded into an $L$-level SID via RQ-VAE, with each level drawn from a
codebook of $K$ codes. A user's engagement history is represented as a
sequence of items, each paired with text metadata (e.g., title or description)
and its SID. We extend the LLM's tokenizer with $L \times K$ hierarchical SID
tokens so the model can both consume and produce SIDs natively.

We train with causal language modeling loss on three types of sequences,
sampled randomly during training:
\begin{itemize}
    \item \textbf{Interleaved sequence}: $(\text{text}_1, \text{sid}_1,
    \text{text}_2, \text{sid}_2, \ldots)$ or with SID before text.
    \item \textbf{Single pair}: $(\text{text}_i, \text{sid}_i)$ or
    $(\text{sid}_i, \text{text}_i)$ for one item.
    \item \textbf{SID-only sequence}: $(\text{sid}_1, \text{sid}_2,
    \text{sid}_3, \ldots)$ without text.
\end{itemize}
This single-stage multi-task training is simpler than the multi-stage pipelines
used by prior work~\cite{he2026plum,liu2025onerec}, but captures the same
core capabilities: sequential prediction, item-text alignment, and SID
understanding.

\subsection{SID Prefix Memory}
\label{sec:method:prefix_memory}

The problem with the setup above is that each SID level code gets a single
learned embedding, regardless of what precedes it. Yet the same code at level
$\ell$ can mean very different things depending on the prefix: there are
$K^{\ell-1}$ possible contexts, each potentially representing a distinct item
cluster. The LLM must learn not only what each code means in context, but also
which prefix-code combinations are valid and which are popular among users.
This distribution is highly skewed, making the learning problem both large and
sparse.

\begin{figure}[t]
  \centering
  \includegraphics[width=\linewidth]{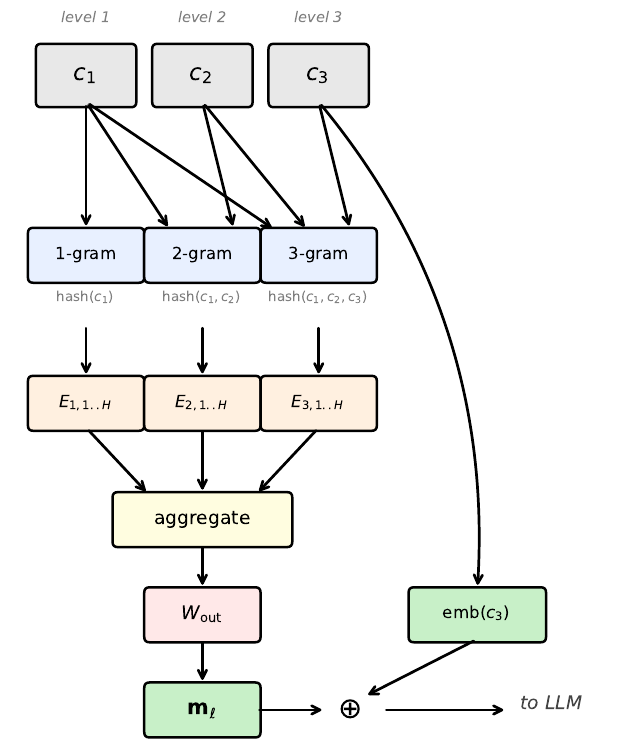}
  \caption{The \ours{} module at level $\ell{=}3$. Given prefix
  $(c_1, c_2, c_3)$, prefix $n$-grams of increasing length are hashed into
  multi-head embedding tables. Retrieved vectors are aggregated, projected via
  $W_{\text{out}}$, and added ($\oplus$) to the input embedding of $c_3$,
  enriching it before the LLM predicts $c_4$.}
  \label{fig:prefixmem}
\end{figure}

\ours{} addresses this directly (Figure~\ref{fig:prefixmem}). Before each SID
token enters the transformer, the encoder hashes the preceding levels into a
compact prefix-conditioned vector and adds it to the token embedding. The
result is that the LLM sees a different input representation for the same code
depending on what came before it.

Concretely, for a SID span $(c_1, \ldots, c_L)$, the encoder computes a
prefix embedding at each level $\ell$ by hashing prefix $n$-grams of increasing
length. Given prefix $(c_1, \ldots, c_\ell)$, for each $n$-gram order
$n \in \{1, \ldots, \min(\ell, N_{\max})\}$ and each hash head
$h \in \{1, \ldots, H\}$:

\begin{equation}
    \text{idx}_{n,h} = \left(\bigoplus_{i=1}^{n} (c_{i} \times p_{i,h})\right) \bmod T
\end{equation}

\noindent where $\times$ is integer multiplication, $\oplus$ is bitwise XOR,
$p_{i,h}$ are fixed prime constants, and $T$ is the hash table size.
The $n$-gram of order $n$ hashes the first $n$ levels of the prefix:
$(c_1)$, $(c_1, c_2)$, $\ldots$, $(c_1, \ldots, c_\ell)$. The retrieved
embeddings from table $E_{n,h} \in \mathbb{R}^{T \times d/H}$ are
concatenated across heads, summed across $n$-gram orders, and projected:

\begin{equation}
    \mathbf{m}_\ell = W_{\text{out}} \sum_{n=1}^{\min(\ell, N_{\max})}
    \text{concat}\left[E_{n,h}[\text{idx}_{n,h}]\right]_{h=1}^{H}
\end{equation}

\noindent The output $\mathbf{m}_\ell \in \mathbb{R}^{d_{\text{model}}}$ is
added to the input embedding of $c_\ell$, enriching it so the LLM can better
predict $c_{\ell+1}$. $W_{\text{out}}$ is initialized near zero so the
encoder's contribution is negligible at initialization, allowing the LLM to
train undisturbed in early steps while the tables warm up.

\subsection{Pre-Training the SID Encoder}
\label{sec:method:pretrain}

Like vision encoders that are pre-trained on image data before being connected
to an LLM, the SID encoder can be pre-trained as well. Even without
pre-training, the encoder improves results substantially
(Section~\ref{sec:results}), but pre-trained embedding tables can further
boost performance. In particular, the distributional structure of Semantic IDs
 can be captured
with simple and inexpensive methods.

We explore a ladder of pre-training sources, each adding one layer of
information over the previous:
\begin{enumerate}
    \item \textbf{Classification head}: Predict $c_{\ell+1}$ from
    $\mathbf{m}_\ell$ via a linear head and cross-entropy loss. This embeds
    prefix-to-next-level transition statistics into the tables, capturing which
    codes tend to follow which prefixes.
    \item \textbf{Generative retrieval model}: Attach the encoder to a
    generative SID model (e.g., Tiger~\cite{rajput2023recommender}) that
    produces SID sequences from user histories. The tables now encode
    behavioral SID generation patterns learned from user interactions.
    \item \textbf{Small LLM}: Train the encoder jointly with a smaller LLM
    (e.g., Qwen3 0.6B) on text+SID data. This infuses language-grounded
    information into the tables, since the encoder must produce representations
    that a transformer can use for both SID prediction and text generation.
\end{enumerate}

After pre-training, the embedding tables are loaded into the target LLM with
a freshly initialized projection layer $W_{\text{out}}$.

\section{Experimental Setup }
\label{sec:experiments}

\subsection{Dataset}
We sample a small portion of Pinterest's user engagement data for
training. The sampled dataset contains approximately 10M subsequences
(each up to 32 engagement events), totaling approximately 240M item
occurrences covering tens of millions of unique items. Each item is encoded into a 5-level SID (2048 codes per
level) by applying RQ-VAE to multimodal embeddings from
PinCLIP~\cite{pinclip2026}; the codebook is trained over the full
billion-item catalog. Text features are image descriptions
generated by a vision-language model. Table~\ref{tab:data} shows
examples of SID-text pairs.

Training and evaluation are split by both time and users: evaluation
sequences come from a later date range with a disjoint set of users,
ensuring no data leakage between the two sets.

\begin{table}[t]
\caption{Examples of item SIDs and their text annotations.}
\label{tab:data}
\begin{tabular}{p{0.28\linewidth} p{0.68\linewidth}}
\toprule
\textbf{SID} & \textbf{Text} \\
\midrule
\texttt{1273, 505, 1934, 1882, 1288} &
This image showcases a sneaker with a beige and brown color scheme. The shoe features a prominent Nike swoosh on the side, a lace-up front, and a thick sole. \\
\midrule
\texttt{1331, 349, 1265, 1247, 399} &
This image showcases a yellow T-shirt with the text ``I LOVE MY HUSBAND BUT SOMETIMES I WANNA SQUARE UP'' printed in bold black letters on the front. \\
\midrule
\texttt{974, 1170, 182, 1140, 865} &
This image showcases a split design featuring a vintage Walkie-Talkie on the left and a woman in a white dress and black hat on the right, with a blue question mark between them. \\
\bottomrule
\end{tabular}
\end{table}

\subsection{Models and Training}
Our default LLM is Qwen3~1.7B~\cite{yang2025qwen3}. To test
generalization, we also train Qwen3~0.6B, Qwen3~4B,
Llama~3.2~1B~\cite{grattafiori2024llama}, and
Gemma~3~1B~\cite{kamath2025gemma}. We selected the learning rate via a sweep over $10^{-3}$ to $10^{-5}$
on Qwen3~1.7B and used the same setting for all other models to ensure
a controlled comparison (the with/without encoder comparison on each
model uses identical hyperparameters). All models are trained for 50K
steps using the multi-task setup described in
Section~\ref{sec:method:setup}.

The \ours{} encoder uses $H$=4 hash heads, default table size $T$=2M
(ablated in Section~\ref{sec:table_size}), memory dimension
$d_{\text{mem}}$=256, and max $n$-gram order $N_{\max}$=4. The encoder
is activated at levels~4 and~5, where the prefix is 3--4 codes long and
the combinatorial space is large enough to benefit from dedicated memory
(see Section~\ref{sec:table_size} for the full per-level analysis). The
encoder's learning rate is set to 5$\times$ the LLM learning rate, since
the sparse hash tables require more aggressive updates to warm up.

We compare the following configurations:
\begin{itemize}
    \item \textbf{Baseline}: LLM trained on mixed text+SID sequences
    without any SID encoder.
    \item \textbf{+ \ours{} (random init)}: Same training with \ours{}
    attached, hash tables initialized randomly.
    \item \textbf{+ \ours{} (pretrained)}: Hash tables pre-trained via
    classification (Section~\ref{sec:method:pretrain}), then loaded for
    joint LLM training.
    \item \textbf{SID-Transformer}: A 4-layer causal transformer encoder
    over the SID prefix, as an alternative encoder architecture.
\end{itemize}

\subsection{Evaluation Protocol}
\label{sec:eval_protocol}

All models are evaluated on 100K held-out examples. For each example,
we sample a random position in a user's engagement sequence and use the
preceding events as history to predict the SID at the sampled position.

We distinguish two types of SID accuracy throughout the paper:
\begin{itemize}
    \item \textbf{Per-level accuracy}: whether the prediction at level
    $\ell$ alone is correct, regardless of other levels. This isolates
    the model's knowledge at each level independently.
    \item \textbf{Prefix accuracy}: whether \emph{all} levels
    $1, \ldots, \ell$ are correct simultaneously. A 5-level prefix match
    means the model produced the correct SID (all 5 codes correct),
    though multiple items may share the same SID due to codebook
    collisions.
\end{itemize}

\noindent We report three families of metrics:

\paragraph{Teacher-forcing per-level accuracy (TF-L$\ell$).}
Our primary metric. Given the user's history and the target item's
ground-truth SID tokens as input, we run a single forward pass and check
whether the model's top-1 prediction at each level position of the
target SID matches the ground truth. This is a \emph{per-level} metric:
TF-L5 measures whether the model predicts the correct 5th code, given
the history context and correct codes at levels~1--4 as input. Because
teacher forcing supplies the correct prefix at every level, it isolates
the model's per-level knowledge without autoregressive error propagation.
In autoregressive generation, an error at level~2 cascades to all deeper
levels, making it impossible to assess whether the model \emph{knows}
the right code at level~5 even if it cannot reach it via greedy decoding.

\paragraph{Full-SID Recall@$K$.}
Using beam search with $K$ beams, we generate SIDs autoregressively
(5 levels sequentially, maintaining $K$ candidates ranked by cumulative
log-probability) and check whether the complete 5-level ground-truth SID
appears anywhere in the top-$K$ candidate set. All 5 codes must match in
a single beam candidate for the item to be retrieved. We report results
for $K \in \{20, 30, 50, 100\}$.

\paragraph{BLEU (SID$\to$text grounding).}
Given a SID as input, the model generates a text description, which is
compared to the reference via BLEU score. This measures how well the
model has learned to ground Semantic IDs in natural language.

\paragraph{SID hit rate.}
For each predicted SID prefix at level~$\ell$, we check whether the
corresponding bucket in the catalog lookup table contains at least one
real item. This measures whether the model produces \emph{retrievable}
SIDs rather than hallucinating non-existent codes.

\section{Results}
\label{sec:results}

We organize results around four questions: (1)~how much does the encoder
improve per-level accuracy and how does this translate to retrieval
(Sections~\ref{sec:table_size}--\ref{sec:recall}), (2)~on which examples
does the gain concentrate (Section~\ref{sec:where}), (3)~does the
approach generalize across model sizes, families, and non-LLM
architectures (Section~\ref{sec:generalization}), and (4)~what encoder
design properties matter (Section~\ref{sec:alternatives}).
Unless otherwise noted, we report TF-L5 as the primary accuracy metric
since the encoder's effect concentrates at the deepest levels.

\subsection{Per-Level Accuracy and Convergence}
\label{sec:table_size}

\begin{table}[t]
\caption{Teacher-forcing accuracy (\%) on Qwen3~1.7B at 50K steps.
The encoder is active at L4 and L5; L1--L3 are unchanged
(${\sim}$42/29/29\%) and omitted for space.}
\label{tab:main}
\begin{tabular}{lcc}
\toprule
 Method & \textbf{TF-L4} & \textbf{TF-L5} \\
\midrule
Baseline          & 33.3 & 37.6 \\
+ \ours{} (500K)  & 40.1 & 49.7 \\
+ \ours{} (2M)    & 42.6 & 54.8 \\
+ \ours{} (5M)    & \textbf{43.4} & \textbf{57.2} \\
\bottomrule
\end{tabular}
\end{table}

Table~\ref{tab:main} shows teacher-forcing accuracy at the SID
levels at matched training compute (50K steps). The results reveal a
clear pattern: levels~1--3 are essentially unchanged by the encoder,
while levels~4 and~5 improve dramatically --- +28\% relative at L4 and
+46\% relative at L5 with the default 2M table. This aligns with the
structure of the problem: at level~1 there is no prefix to condition on;
at level~2 the prefix is a single code (2048 possible values); but by
level~4 the prefix is 3 codes, and the number of \emph{observed} prefix
patterns in the training corpus grows rapidly. The hash tables provide
dedicated capacity for storing these prefix-to-code relationships, while
the LLM must encode them implicitly in its shared parameters.

Table~\ref{tab:main} also serves as a table-size ablation. Scaling from
500K to 2M rows yields +10\% relative at L5 (49.7\% $\to$ 54.8\%), while
2M to 5M yields only +4\% (54.8\% $\to$ 57.2\%), indicating
diminishing returns. We use 2M as the default for subsequent experiments
unless otherwise noted.

We also evaluate a configuration that activates the encoder at L3--L5
(2M table, requiring at least a 2-code prefix). The gain scales with
prefix length: L3 improves modestly (+6\% relative, from 29.2\% to
30.9\%), while L4 improves +27\% relative (34.5\%$\to$43.9\%) and L5
improves +46\% relative (37.6\%$\to$54.8\%), where the prefix space
grows exponentially ($2048^3$ to $2048^4$ possible contexts). Based on
this, we activate the encoder at L4 and L5 only in subsequent
experiments, where the prefix space is large enough to benefit
significantly from dedicated memory.

\begin{figure}[t]
  \centering
  \includegraphics[width=\linewidth]{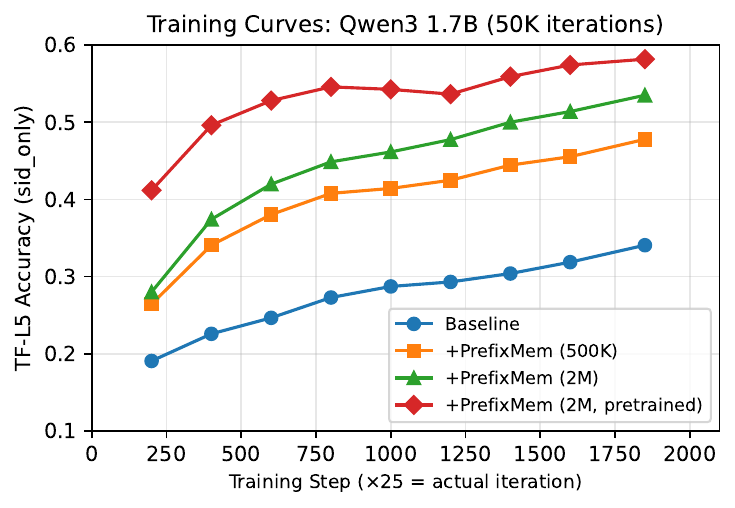}
  \caption{TF-L5 accuracy over training on Qwen3~1.7B. The encoder's
  advantage emerges early and widens throughout; pre-trained tables
  provide an immediate head start.}
  \label{fig:convergence}
\end{figure}

Figure~\ref{fig:convergence} confirms that the accuracy gap is not a
convergence artifact. The encoder pulls ahead within the first few
thousand iterations and the gap widens throughout training. The baseline
plateaus around 35\% while the 2M encoder reaches 53\% and the
pretrained encoder reaches 58\%. This widening gap indicates the encoder
provides dedicated capacity for prefix patterns that the LLM alone
struggles to acquire at matched compute.

\subsection{Encoder Pre-Training and Transfer}
\label{sec:pretrain_results}

A key advantage of treating the SID encoder as a separate module is that
it can be pre-trained independently --- like a vision encoder trained on
images before being connected to an LLM. We evaluate several
initialization strategies, all followed by the same 50K steps of joint
LLM training on Qwen3~1.7B.

\begin{table}[t]
\caption{Effect of encoder initialization on Qwen3~1.7B after 50K joint
training steps. All rows use the same LLM and training data; only the
encoder's starting point differs.}
\label{tab:warmstart}
\begin{tabular}{lcccc}
\toprule
\textbf{Initialization} & \textbf{TF-L4} & \textbf{TF-L5} & \textbf{R@100} & \textbf{BLEU} \\
\midrule
Random init (2M)          & 42.6 & 54.8 & 11.3 & 29.1 \\
Random init (5M)          & 43.4 & 57.2 & 11.6 & 32.3 \\
Cls.\ pretrain (2M)       & 43.2 & 58.5 & 11.2 & 26.3 \\
Cls.\ pretrain (5M)       & \textbf{49.6} & \textbf{64.0} & 11.8 & 27.2 \\
LLM pretrain (2M, 1.7B)       & 45.3 & 57.0 & 11.9 & \textbf{33.1} \\
LLM pretrain (2M, 0.6B)       & 45.3 & 56.8 & 11.8 & \textbf{33.1} \\
Tiger pretrain (2M)       & 46.7 & 61.2 & \textbf{12.3} & 28.2 \\
\bottomrule
\end{tabular}
\end{table}

Table~\ref{tab:warmstart} shows the results. Classification
pre-training --- a simple objective that predicts $c_{\ell+1}$ from the
prefix embedding $\mathbf{m}_\ell$ via a linear head and cross-entropy
loss --- with 5M tables achieves the highest TF-L5 of 64.0\%, a +70\%
relative gain over the no-encoder baseline. This pre-training is
inexpensive: it involves only sparse table lookups and a linear
classifier, with no transformer forward passes.

Two additional strategies confirm that the encoder's learned
representations are portable:
\begin{itemize}
    \item \textbf{LLM pretrain (1.7B)}: Loading hash tables from a prior
    run where the encoder was trained jointly with Qwen3~1.7B (with a
    freshly initialized $W_{\text{out}}$) yields TF-L5 of 57.0\%,
    outperforming random initialization (54.8\%).
    \item \textbf{LLM pretrain (0.6B)}: Loading tables from a Qwen3~0.6B
    run into the 1.7B model (re-initializing $W_{\text{out}}$ to match
    the larger hidden dimension) yields TF-L5 of 56.8\%. The hash tables
    are model-agnostic; only the output projection needs to match the
    target LLM.
\end{itemize}

The BLEU column reveals an interesting dissociation: LLM-pretrained tables
achieve much higher BLEU (33.1) than classification-pretrained tables
(26--27), despite the latter having higher TF-L5 (64.0 vs.\ 57.0).
Classification pre-training optimizes for next-code prediction, which
directly improves TF; LLM pre-training optimizes for representations that
a transformer can use for both SID prediction \emph{and} text generation,
which benefits BLEU. This mirrors the distinction between task-specific
and general-purpose pre-training in vision encoders.

The Tiger row shows another pre-training source: hash tables trained
jointly with Tiger~\cite{rajput2023recommender} (a generative retrieval
model), then loaded into Qwen3~1.7B for joint fine-tuning. At the same
2M table size, Tiger-pretrained tables achieve the highest Recall@100
(12.3\%) and strong TF-L5 (61.2\%), outperforming both classification
and LLM pre-training. We attribute this to Tiger's training objective:
it learns to generate SIDs that users actually engage with next, directly
encoding user-behavioral transition patterns into the hash tables. In
contrast, classification pre-training only captures static co-occurrence
statistics, and LLM pre-training splits capacity between SID and language
tasks.

In subsequent sections, ``pretrained 5M'' refers to the classification
pre-trained encoder with 5M tables (the strongest TF configuration).

\subsection{Retrieval and Catalog Validity}
\label{sec:recall}

Teacher-forcing accuracy measures per-level knowledge in isolation. For
deployment, what matters is whether the model can retrieve the correct
item via autoregressive beam search, and whether its predictions
correspond to real catalog entries.

\begin{table}[t]
\caption{Full-SID Recall@$K$ (\%): fraction of examples where the exact
5-level ground-truth SID appears in the top-$K$ beam candidates.}
\label{tab:recall}
\begin{tabular}{lcccc}
\toprule
\textbf{Method} & \textbf{R@20} & \textbf{R@30} & \textbf{R@50} & \textbf{R@100} \\
\midrule
Baseline                   & 6.0 & 6.9 & 8.0 &  9.5 \\
+ \ours{} (2M)             & 6.6 & 7.7 & 9.2 & 11.3 \\
+ \ours{} (5M)             & \textbf{6.7} & \textbf{7.8} & \textbf{9.4} & 11.6 \\
+ \ours{} (pt.\ 5M)       & 6.3 & 7.5 & 9.3 & \textbf{11.8} \\
\midrule
$\Delta$ (5M vs.\ base)    & +11\% & +14\% & +18\% & +22\% \\
\bottomrule
\end{tabular}
\end{table}

\paragraph{Recall scales with beam width.}
Table~\ref{tab:recall} shows that the encoder's retrieval benefit grows
monotonically with beam width: +11\% relative at $K$=20, reaching +22\%
at $K$=100. This monotonic scaling occurs because wider beams explore
more prefix paths, allowing the encoder's improved deep-level predictions
to surface in the candidate set.

This property is particularly relevant for production recommender systems,
where the generative model typically produces a large candidate pool
(hundreds or thousands of items) that is subsequently re-ranked by a
scoring model. In such systems, the encoder's value increases with the
retrieval budget.

\begin{table}[t]
\caption{Beam search diversity: average number of distinct L1 clusters
among the top-100 beam candidates.}
\label{tab:diversity}
\begin{tabular}{lc}
\toprule
\textbf{Method} & \textbf{Unique L1 clusters} \\
\midrule
Baseline              & 5.5 \\
+ \ours{} (2M)        & 6.0 \\
+ \ours{} (5M)        & 6.0 \\
+ \ours{} (pt.\ 5M)  & \textbf{6.2} \\
\bottomrule
\end{tabular}
\end{table}

\paragraph{How the encoder reshapes beam search.}
\label{sec:beam_diversity}
Table~\ref{tab:diversity} shows that the encoder increases beam
diversity from 5.5 to 6.2 unique L1 clusters (+13\%). The encoder's
sharper L4/L5 predictions modify cumulative beam scores, causing
different L1 prefixes to survive the ranking. This broader exploration
of the SID space is consistent with the recall improvement: more diverse
beams means more ground-truth prefixes become reachable.

\begin{table}[t]
\caption{SID hit rate (\%): fraction of predicted SIDs whose prefix
bucket is non-empty in the catalog. Higher = fewer hallucinated SIDs.}
\label{tab:dbhit}
\begin{tabular}{lccccc}
\toprule
\textbf{Method} & \textbf{L1} & \textbf{L2} & \textbf{L3} & \textbf{L4} & \textbf{L5} \\
\midrule
Baseline               & 100 & 99.8 & 95.7 & 76.5 & 50.7 \\
+ \ours{} (500K)       & 100 & 99.9 & 96.4 & 78.3 & 55.8 \\
+ \ours{} (2M)         & 100 & 99.9 & 96.5 & 78.6 & 57.6 \\
+ \ours{} (5M)         & 100 & 99.8 & 96.1 & 78.9 & 58.9 \\
Cls.\ pretrain (2M)    & 100 & 100.0 & 97.0 & 81.6 & 62.8 \\
Cls.\ pretrain (5M)    & 100 & 99.9 & 96.8 & \textbf{82.4} & \textbf{65.1} \\
\bottomrule
\end{tabular}
\end{table}

\paragraph{Catalog validity.}
\label{sec:db_hit}
Beyond retrieving the \emph{correct} SID, a practical question is whether
predictions correspond to \emph{real} catalog items. The SID space is
highly sparse at deep levels ($2048^5$ possible combinations vs.\ tens of
millions of items observed in training), so most incorrect predictions
land in empty
buckets. This matters in recommendation systems: invalid SIDs will be discarded before serving, effectively halving usable candidates and doubling the beam-search compute needed to fill the candidate budget. Table~\ref{tab:dbhit} shows our encoder improves L5 hit rate
from 50.7\% to 65.1\% (+28\% relative): the baseline hallucinates
non-existent SIDs for nearly half of predictions, while the pretrained
encoder reduces this to one-third. A similar ${\sim}$50\% hit rate was observed by Deng et
al.~\cite{deng2025onerec}, who addressed it via GRPO-style
reinforcement learning~\cite{shao2024deepseekmath} to improve SID
validity. However, recent work~\cite{chen2026does} suggests that
RL-trained models remain bounded by the base model's underlying
capabilities. If this translates to our setting, RL can steer generation
toward valid items but may not expand the set of items the model is
capable of producing. Our encoder
operates at a different level: it improves the base model's knowledge of
prefix-to-code transitions, expanding the pool of items the model can
generate correctly. The two approaches are complementary and could be
combined.
\subsection{Where Does the Encoder Help?}
\label{sec:where}

The preceding sections established that the encoder improves accuracy,
recall, and catalog validity. We now analyze \emph{which} examples and
items benefit most.

\paragraph{Hard vs.\ easy examples.}
To understand where the gains concentrate, we introduce a
\emph{conditional} evaluation. During teacher forcing, at each SID level
we rank the 2048 valid codes by the model's logit scores and check
whether the ground-truth code appears in the top~10. We call an example
``reachable'' if the correct code ranks within the top~10 at
\emph{every} prefix level (L1 through L4 simultaneously). If the correct
code falls outside top-10 at \emph{any single} prefix level, the example
is ``unreachable'' --- the model does not consider the correct prefix a
plausible option at some level.

\begin{table}[t]
\caption{TF-L5 accuracy (\%) conditioned on prefix reachability (whether
the ground-truth code ranks top-10 at all prefix levels).}
\label{tab:cond}
\begin{tabular}{lcc}
\toprule
\textbf{Method} & \textbf{Reachable} & \textbf{Unreachable} \\
\midrule
Baseline                   & 43.3 & 36.4 \\
+ \ours{} (2M)             & 44.8 & 54.8 \\
+ \ours{} (5M)             & \textbf{45.1} & 57.3 \\
+ \ours{} (pretrained 5M)  & 44.6 & \textbf{64.5} \\
\midrule
$\Delta$ (pretrained vs.\ base)   & +3.0\% & +77.2\% \\
\bottomrule
\end{tabular}
\end{table}

Table~\ref{tab:cond} reveals a striking asymmetry: on reachable examples
the encoder adds only +3\% relative (43.3\%$\to$44.6\%), but on
unreachable examples it improves accuracy by +77\% relative
(36.4\%$\to$64.5\%). The encoder provides knowledge for prefix patterns
that the LLM cannot rank highly on its own.
This concentration on hard examples explains the recall scaling in
Table~\ref{tab:recall}: wider beams are needed to surface the encoder's
knowledge on difficult prefix patterns.

\paragraph{Popular vs.\ tail items.}
\label{sec:popularity}
We further segment by how often the ground-truth L4 prefix appears in
training data: rare (bottom quartile, mostly singletons), moderate, and
popular (top quartile).

\begin{table}[t]
\caption{TF-L5 accuracy (\%) by L4 prefix popularity. Rare prefixes
appear ${\sim}$1 time in training; popular ones appear many times.}
\label{tab:popularity}
\begin{tabular}{lccc}
\toprule
\textbf{Method} & \textbf{Rare} & \textbf{Medium} & \textbf{Popular} \\
\midrule
Baseline             & 26.8 & 34.8 & 52.5 \\
+ \ours{} (2M)       & 42.9 & 67.9 & 68.9 \\
+ \ours{} (5M)       & 46.4 & 71.1 & 69.5 \\
Cls.\ pretrain (5M)  & \textbf{57.6} & 75.6 & \textbf{70.4} \\
Tiger pretrain (2M)  & 51.9 & \textbf{76.3} & 70.3 \\
\bottomrule
\end{tabular}
\end{table}

Table~\ref{tab:popularity} shows the encoder disproportionately helps
rare items: +115\% relative (26.8\%$\to$57.6\%) vs.\ +34\% relative for popular
items (52.5\%$\to$70.4\%). Popular prefixes have sufficient training
exposure for the LLM to learn their transitions through its transformer
weights. Rare prefixes lack this exposure, but the hash tables learn
prefix-to-code relationships more sample-efficiently: a single O(1)
lookup directly retrieves a prefix-conditioned representation, whereas
the LLM must compose the same knowledge through multiple layers of
attention over the prefix tokens. This efficiency gap is largest for
rare prefixes where the LLM has few training examples to learn from.

\subsection{Generalization}
\label{sec:generalization}

\begin{table}[t]
\caption{TF-L5 (\%) and R@100 (\%) across model families and sizes.
All encoder rows use 2M \ours{} with random initialization.}
\label{tab:scale}
\begin{tabular}{llcc}
\toprule
\textbf{Model} & \textbf{Encoder} & \textbf{TF-L5} & \textbf{R@100} \\
\midrule
Qwen3 0.6B  & ---        & 32.1 &  8.4 \\
Qwen3 0.6B  & \ours{}    & 54.1 & 10.5 \\
\midrule
Qwen3 1.7B  & ---        & 37.6 &  9.5 \\
Qwen3 1.7B  & \ours{}    & 54.8 & 11.3 \\
\midrule
Qwen3 4B    & ---        & 45.0 & 10.9 \\
Qwen3 4B    & \ours{}    & \textbf{55.8} & \textbf{12.0} \\
\midrule
Llama 3.2 1B & ---       & 39.2 &  9.5 \\
Llama 3.2 1B & \ours{}   & 54.5 & 11.1 \\
\midrule
Gemma 3 1B   & ---       & 37.3 &  9.3 \\
Gemma 3 1B   & \ours{}   & 54.4 & 11.1 \\
\bottomrule
\end{tabular}
\end{table}

\paragraph{Across LLM families and scales.}
Table~\ref{tab:scale} evaluates \ours{} across three Qwen model sizes
and two additional LLM families. Three findings stand out:

\textbf{Consistent gains across scales.} \ours{} improves TF-L5 by
+69\% relative on the 0.6B model, +46\% on 1.7B, and +24\% on 4B. The
relative gain is largest on the smallest model, consistent with the
hypothesis that the encoder provides dedicated capacity for SID prefix
patterns that smaller models cannot fit in their shared parameters.

\textbf{Encoder outweighs model scaling.} Qwen3~0.6B \emph{with}
\ours{} achieves TF-L5 of 54.1\%, surpassing Qwen3~4B \emph{without}
it (45.0\%). Under our evaluation protocol, a 2M-row hash table
provides larger gains for SID prediction than an 8$\times$ increase in
LLM parameters. This suggests that adding the encoder to a small LLM
may be preferable to scaling up the model, both for training efficiency
(fewer FLOPs per step) and serving cost (smaller model to deploy).

\textbf{Architecture-agnostic.} Llama~3.2~1B and Gemma~3~1B without
encoder achieve TF-L5 of 39.2\% and 37.3\%; with \ours{} they reach
54.5\% and 54.4\% (+39\% and +46\% relative), comparable to
Qwen3~1.7B (54.8\%). The encoder is not tied to a specific LLM
architecture and functions as a modality-specific module that can be
plugged into any transformer backbone.

\begin{table}[t]
\caption{Tiger (35M-param generative retrieval model) with and without
\ours{}. TF accuracy and Recall@10 (beam search with 10 beams).}
\label{tab:tiger}
\begin{tabular}{lccc}
\toprule
\textbf{Encoder} & \textbf{TF-L4} & \textbf{TF-L5} & \textbf{R@10 L5} \\
\midrule
None     & 27.6 & 27.4 & 1.01 \\
\ours{}  & \textbf{46.3} & \textbf{60.6} & \textbf{1.35} \\
\midrule
$\Delta$ & +68\% & +121\% & +34\% \\
\bottomrule
\end{tabular}
\end{table}

\paragraph{Beyond LLMs.}
\label{sec:tiger}
\ours{} is not specific to LLMs --- it can augment any autoregressive SID
generator. Table~\ref{tab:tiger} shows results on
Tiger~\cite{rajput2023recommender}, a 35M-parameter generative retrieval
model. Without the encoder, Tiger's L4 and L5 accuracy are both
${\sim}$27\%, indicating it cannot learn deep-level structure with its
limited parameters. With \ours{}, L5 reaches 60.6\% (+121\% relative)
and Recall@10 improves by +34\%, confirming the encoder benefits any
model that generates SID tokens hierarchically.

\subsection{Encoder Architecture and Design}
\label{sec:alternatives}

\begin{table}[t]
\caption{SID encoder architecture comparison on Qwen3~1.7B at 50K steps.}
\label{tab:alternatives}
\begin{tabular}{lcc}
\toprule
\textbf{Method} & \textbf{TF-L5} & \textbf{BLEU} \\
\midrule
Baseline (no encoder)        & 37.6 & 23.3 \\
+ \ours{} (hash memory)      & \textbf{54.8} & \textbf{30.9} \\
+ SID-Transformer (4-layer)  & 39.6 & 23.4 \\
\bottomrule
\end{tabular}
\end{table}

To understand why hash-based memory works better than learned
computation, we compare \ours{} against an alternative encoder: a
4-layer causal transformer (336M parameters) that processes per-level embeddings of the
SID prefix via self-attention, producing a contextualized prefix
representation that is added to the token embedding (the same injection
point as \ours{}).

Table~\ref{tab:alternatives} shows a stark contrast. The
SID-Transformer provides only marginal L5 improvement (+5\% relative)
and no BLEU gain, while \ours{} achieves +46\% at L5 and +33\% on BLEU.
Despite having 336M dense parameters, the transformer
cannot memorize the combinatorial prefix space through learned
computation alone. In contrast, \ours{} stores ${\sim}$2B parameters in
sparse hash tables where each prefix pattern maps directly to a
dedicated entry via O(1) lookup --- no gradient-based learning of
prefix compositions required. The SID encoding problem is fundamentally
about \emph{memory capacity} (storing prefix-conditioned information)
rather than \emph{learned computation} (composing representations
through attention). More broadly, encoder architecture design for
hierarchical codes remains an open question: how to balance memory
capacity, computational cost, and generalization beyond hash tables
(e.g., learned routing, mixture-of-experts, or hybrid approaches) is a
direction for future work.

\paragraph{Design choices within \ours{}.}
Given that hash-based memory is effective, we ablate two internal design
choices in Table~\ref{tab:ablation}.

\begin{table}[t]
\caption{Encoder design choices on Qwen3~1.7B with 2M tables.}
\label{tab:ablation}
\begin{tabular}{llccc}
\toprule
\textbf{N-gram scale} & \textbf{Slice direction} & \textbf{TF-L5} & \textbf{R@100} & \textbf{BLEU} \\
\midrule
Multi-scale   & prefix  & 54.8 & 11.3 & 30.9 \\
Multi-scale   & suffix  & 54.6 & 11.4 & 29.9 \\
Single-scale  & N/A              & 54.0 & 11.4 & 30.4 \\
\bottomrule
\end{tabular}
\end{table}

\begin{itemize}
    \item \textbf{N-gram scale}: ``Multi-scale'' hashes prefix $n$-grams
    of length 1 through $\min(\ell, N_{\max})$, providing representations
    at multiple granularities. ``Single-scale'' hashes only the complete
    prefix as one unit.
    \item \textbf{Slice direction} (multi-scale only): When extracting a
    partial $n$-gram of length $n < \ell$, ``suffix'' takes the last $n$
    codes (most recent); ``prefix'' takes the first $n$
    (position-anchored). Single-scale always uses the full prefix so this
    choice does not apply.
\end{itemize}

All three configurations achieve similar performance: TF-L5 ranges from
54.0\% to 54.8\%, R@100 from 11.3\% to 11.4\%, and BLEU from 29.9 to
30.9. The encoder is robust to these design choices. Multi-scale with
All configurations perform comparably, and the
differences are small enough that the choice is not critical.

\section{Discussion }
\label{sec:discussion}

\paragraph{Overhead Analysis.}
With default settings ($T$=2M, $H$=4, $N_{\max}$=4, $d_{\text{mem}}$=256),
\ours{} contains ${\sim}$2B parameters in embedding tables. While large
in raw count, these differ fundamentally from transformer parameters:

\textbf{Compute.}
For each active SID position (2 out of 5 levels by default), the encoder
performs $H \times N_{\max} = 16$ table lookups, one summation across
$n$-gram orders, and one output projection ($d_{\text{mem}} \to
d_{\text{model}}$: $256 \times 2048 \approx 0.5$M multiply-adds). Per
5-level SID span this totals ${\sim}$2M FLOPs, vs.\ ${\sim}$17B for the
LLM forward pass on the same 5 tokens (using the standard ${\approx}2P$
FLOPs-per-token estimate~\cite{kaplan2020scaling}). The encoder thus adds
$<$0.02\% compute overhead. Empirically, we observe $<$3\% training
throughput decrease at matched batch size.

\textbf{Memory.}
The embedding tables and their AdamW optimizer states add ${\sim}$10~GB
peak GPU memory (measured as the increase in max reserved memory with
vs.\ without the encoder on the same hardware). Smaller tables (500K
rows) reduce this 4$\times$ with modest accuracy loss
(Table~\ref{tab:main}).

\textbf{Serving.}
At inference, the encoder adds one embedding lookup and one small
projection per generated SID token --- no attention, no layer
normalization, no sequential dependence. Its cost is negligible relative
to the LLM's autoregressive decoding. Since only $O(1)$ rows are accessed per token, mature embedding
infrastructure optimizations (CPU offloading, quantization,
mixed-dimension tables) can be directly applied to further reduce memory
and serving cost or to further scale up the embedding tables.

\paragraph{Encoder Pre-Training as an Inductive Bias.}
How to pre-train the encoder is an interesting open question. Different
pre-training strategies infuse different inductive biases into the final
generative recommender: classification pre-training encodes static
transition statistics (highest TF accuracy), LLM pre-training encodes
language-grounded representations (highest BLEU), and Tiger pre-training
encodes behavioral generation patterns (highest recall). Similarly, the
training corpus biases which SID transitions the tables encode ---
engagement-weighted data favors exploitation of known preferences, while
catalog-uniform data would favor diversity and cold-start coverage. We
believe exploring these pre-training strategies and their downstream
effects in real traffic settings is a promising direction for future
research.

\paragraph{Language Capability Tradeoff.}
A known challenge in adapting LLMs for SID-based generative
recommendation is the tradeoff between acquiring new SID knowledge and
retaining pre-trained language capabilities. Verma et al.~\cite{verma2026orbit} proposed model merging
to mitigate this but still observed degradation in one or both
capabilities. Our encoder offers a potentially better tradeoff: since SID prefix
knowledge is offloaded to external hash tables, the LLM's transformer
weights bear less burden for memorizing prefix patterns. At matched
training iterations, the encoder achieves substantially better SID
accuracy while applying the same number of gradient updates to the LLM,
meaning language capability degradation should be comparable to the
baseline. Alternatively, the encoder can reach the baseline's SID
accuracy in fewer training steps (Figure~\ref{fig:convergence}),
reducing the total exposure to catastrophic forgetting.

\paragraph{Limitations.}
The encoder's benefit depends on learnable structure in the SID prefix
space. If RQ-VAE codebook transitions are near-uniform (e.g., from heavy
regularization), the tables have less to learn. The hash table size must
also be scaled to the effective number of distinct prefix patterns.
Finally, the encoder requires a non-trivial prefix to be useful. At
level~1 there is no prefix; at level~2 the prefix is a single code from
a vocabulary of 2048, which the LLM can memorize directly. The encoder
becomes effective starting at level~3 (2-code prefix with ${\sim}$4M
combinations) and provides its largest gains at levels~4--5 where the
prefix space grows into the millions. For SID hierarchies with fewer
than 3 levels, the encoder would provide little benefit.

\section{Conclusion }
\label{sec:conclusion}

We showed that Semantic IDs in generative recommendation benefit from a
dedicated encoder, just as images and audio do in multimodal LLMs. Our encoder,
\ours{}, is a lightweight prefix memory module that provides structured,
prefix-conditioned representations at SID token positions. It improves
deepest-level SID accuracy by 46\% relative and retrieval recall by 22\%,
with gains concentrating on hard examples where greedy decoding fails (up to
+77\% relative) and on rare items that lack sufficient training exposure
(+115\% relative). The encoder can be pre-trained cheaply and transferred
across LLM families (Qwen, Llama, Gemma) and non-LLM models (Tiger), and a
0.6B model with \ours{} outperforms a 4B model without it. By decoupling SID
knowledge from the LLM's parameters, \ours{} offers a practical path toward
making LLM-based generative recommendation more affordable and scalable. We
hope this perspective of treating SIDs as a first-class modality informs
future work on efficient generative recommendation systems.


\bibliographystyle{ACM-Reference-Format}
\bibliography{references}

@article{rajput2023recommender,
  title={Recommender systems with generative retrieval},
  author={Rajput, Shashank and Mehta, Nikhil and Singh, Anima and Hulikal Keshavan, Raghunandan and Vu, Trung and Heldt, Lukasz and Hong, Lichan and Tay, Yi and Tran, Vinh and Samost, Jonah and others},
  journal={Advances in Neural Information Processing Systems},
  volume={36},
  pages={10299--10315},
  year={2023}
}

@inproceedings{singh2024better,
  title={Better generalization with semantic ids: A case study in ranking for recommendations},
  author={Singh, Anima and Vu, Trung and Mehta, Nikhil and Keshavan, Raghunandan and Sathiamoorthy, Maheswaran and Zheng, Yilin and Hong, Lichan and Heldt, Lukasz and Wei, Li and Tandon, Devansh and others},
  booktitle={Proceedings of the 18th ACM Conference on Recommender Systems},
  pages={1039--1044},
  year={2024}
}

@inproceedings{he2026plum,
  title={Plum: Adapting pre-trained language models for industrial-scale generative recommendations},
  author={He, Ruining and Heldt, Lukasz and Hong, Lichan and Keshavan, Raghunandan and Mao, Shifan and Mehta, Nikhil and Su, Zhengyang and Tsai, Alicia and Wang, Yueqi and Wang, Shao-Chuan and others},
  booktitle={Proceedings of the ACM Web Conference 2026},
  pages={8093--8104},
  year={2026}
}

@article{deng2025onerec,
  title={Onerec: Unifying retrieve and rank with generative recommender and iterative preference alignment},
  author={Deng, Jiaxin and Wang, Shiyao and Cai, Kuo and Ren, Lejian and Hu, Qigen and Ding, Weifeng and Luo, Qiang and Zhou, Guorui},
  journal={arXiv preprint arXiv:2502.18965},
  year={2025}
}

@article{liu2025onerec,
  title={Onerec-think: In-text reasoning for generative recommendation},
  author={Liu, Zhanyu and Wang, Shiyao and Wang, Xingmei and Zhang, Rongzhou and Deng, Jiaxin and Bao, Honghui and Zhang, Jinghao and Li, Wuchao and Zheng, Pengfei and Wu, Xiangyu and others},
  journal={arXiv preprint arXiv:2510.11639},
  year={2025}
}

@article{cheng2026conditional,
  title={Conditional memory via scalable lookup: A new axis of sparsity for large language models},
  author={Cheng, Xin and Zeng, Wangding and Dai, Damai and Chen, Qinyu and Wang, Bingxuan and Xie, Zhenda and Huang, Kezhao and Yu, Xingkai and Hao, Zhewen and Li, Yukun and others},
  journal={arXiv preprint arXiv:2601.07372},
  year={2026}
}

@inproceedings{zheng2025enhancing,
  title={Enhancing embedding representation stability in recommendation systems with semantic id},
  author={Zheng, Carolina and Huang, Minhui and Pedchenko, Dmitrii and Rangadurai, Kaushik and Wang, Siyu and Xia, Fan and Nahum, Gaby and Lei, Jie and Yang, Yang and Liu, Tao and others},
  booktitle={Proceedings of the Nineteenth ACM Conference on Recommender Systems},
  pages={954--957},
  year={2025}
}

@article{yang2025qwen3,
  title={Qwen3 technical report},
  author={Yang, An and Li, Anfeng and Yang, Baosong and Zhang, Beichen and Hui, Binyuan and Zheng, Bo and Yu, Bowen and Gao, Chang and Huang, Chengen and Lv, Chenxu and others},
  journal={arXiv preprint arXiv:2505.09388},
  year={2025}
}

@inproceedings{lee2022autoregressive,
  title={Autoregressive image generation using residual quantization},
  author={Lee, Doyup and Kim, Chiheon and Kim, Saehoon and Cho, Minsu and Han, Wook-Shin},
  booktitle={Proceedings of the IEEE/CVF conference on computer vision and pattern recognition},
  pages={11523--11532},
  year={2022}
}

@article{liang2026generative,
  title={Generative Reasoning Re-ranker},
  author={Liang, Mingfu and Li, Yufei and Xu, Jay and Asadi, Kavosh and Liu, Xi and Gu, Shuo and Rangadurai, Kaushik and Shyu, Frank and Wang, Shuaiwen and Yang, Song and others},
  journal={arXiv preprint arXiv:2602.07774},
  year={2026}
}

@article{defossez2024moshi,
  title={Moshi: a speech-text foundation model for real-time dialogue},
  author={D{\'e}fossez, Alexandre and Mazar{\'e}, Laurent and Orsini, Manu and Royer, Am{\'e}lie and P{\'e}rez, Patrick and J{\'e}gou, Herv{\'e} and Grave, Edouard and Zeghidour, Neil},
  journal={arXiv preprint arXiv:2410.00037},
  year={2024}
}

@article{liu2023visual,
  title={Visual instruction tuning},
  author={Liu, Haotian and Li, Chunyuan and Wu, Qingyang and Lee, Yong Jae},
  journal={Advances in neural information processing systems},
  volume={36},
  pages={34892--34916},
  year={2023}
}

@inproceedings{letter2024,
  title={Learnable Item Tokenization for Generative Recommendation},
  author={Bao, Honghui and others},
  booktitle={Proceedings of the 33rd ACM International Conference on Information and Knowledge Management (CIKM)},
  year={2024}
}

@inproceedings{cost2024,
  title={Cost: Contrastive quantization based semantic tokenization for generative recommendation},
  author={Zhu, Jieming and Jin, Mengqun and Liu, Qijiong and Qiu, Zexuan and Dong, Zhenhua and Li, Xiu},
  booktitle={Proceedings of the 18th ACM Conference on Recommender Systems},
  pages={969--974},
  year={2024}
}

@inproceedings{eager2024,
  title={Eager: Two-stream generative recommender with behavior-semantic collaboration},
  author={Wang, Ye and Xun, Jiahao and Hong, Minjie and Zhu, Jieming and Jin, Tao and Lin, Wang and Li, Haoyuan and Li, Linjun and Xia, Yan and Zhao, Zhou and others},
  booktitle={Proceedings of the 30th ACM SIGKDD Conference on Knowledge Discovery and Data Mining},
  pages={3245--3254},
  year={2024}
}

@inproceedings{etegrec2025,
  title={Generative recommender with end-to-end learnable item tokenization},
  author={Liu, Enze and Zheng, Bowen and Ling, Cheng and Hu, Lantao and Li, Han and Zhao, Wayne Xin},
  booktitle={Proceedings of the 48th International ACM SIGIR Conference on Research and Development in Information Retrieval},
  pages={729--739},
  year={2025}
}

@article{sidreasoner2026,
  title={Reasoning over Semantic IDs Enhances Generative Recommendation},
  author={He, Yingzhi and Sun, Yan and Tan, Junfei and Chen, Yuxin and Kong, Xiaoyu and Shen, Chunxu and Wang, Xiang and Zhang, An and Chua, Tat-Seng},
  journal={arXiv preprint arXiv:2603.23183},
  year={2026}
}

@article{pinrec2025,
  title={Pinrec: Outcome-conditioned, multi-token generative retrieval for industry-scale recommendation systems},
  author={Agarwal, Prabhat and Badrinath, Anirudhan and Bhasin, Laksh and Yang, Jaewon and Botta, Edoardo and Xu, Jiajing and Rosenberg, Charles},
  journal={arXiv preprint arXiv:2504.10507},
  year={2025}
}

@inproceedings{mtgr2025,
  title={Mtgr: Industrial-scale generative recommendation framework in meituan},
  author={Han, Ruidong and Yin, Bin and Chen, Shangyu and Jiang, He and Jiang, Fei and Li, Xiang and Ma, Chi and Huang, Mincong and Li, Xiaoguang and Jing, Chunzhen and others},
  booktitle={Proceedings of the 34th ACM International Conference on Information and Knowledge Management},
  pages={5731--5738},
  year={2025}
}

@article{snap_sid2026,
  title={Semantic IDs for Recommender Systems at Snapchat},
  title={Semantic IDs for Recommender Systems at Snapchat: Use Cases, Technical Challenges, and Design Choices},
  author={Ju, Clark Mingxuan and Zhao, Tong and Neves, Leonardo and Collins, Liam and Kumar, Bhuvesh and Ren, Jiwen and Zhang, Lili and Zhuo, Wenfeng and Zhang, Vincent and Bai, Xiao and others},
  journal={arXiv preprint arXiv:2604.03949},
  year={2026}
}

@article{zhou2025openonerec,
  title={OpenOneRec Technical Report},
  author={Zhou, Guorui and Bao, Honghui and Huang, Jiaming and Deng, Jiaxin and Zhang, Jinghao and She, Junda and Cai, Kuo and Ren, Lejian and Ren, Lu and Luo, Qiang and others},
  journal={arXiv preprint arXiv:2512.24762},
  year={2025}
}

@article{zhou2025onerec,
  title={OneRec Technical Report},
  author={Zhou, Guorui and Deng, Jiaxin and Zhang, Jinghao and Cai, Kuo and Ren, Lejian and Luo, Qiang and Wang, Qianqian and Hu, Qigen and Huang, Rui and Wang, Shiyao and others},
  journal={arXiv preprint arXiv:2506.13695},
  year={2025}
}

@article{zhou2025onerecv2,
  title={OneRec-v2 Technical Report},
  author={Zhou, Guorui and Hu, Hengrui and Cheng, Hongtao and Wang, Huanjie and Deng, Jiaxin and Zhang, Jinghao and Cai, Kuo and Ren, Lejian and Ren, Lu and Yu, Liao and others},
  journal={arXiv preprint arXiv:2508.20900},
  year={2025}
}

@inproceedings{tay2022transformer,
  title={Transformer memory as a differentiable search index},
  author={Tay, Yi and Tran, Vinh and Dehghani, Mostafa and Ni, Jianmo and Bahri, Dara and Mehta, Harsh and Qin, Zhen and Hui, Kai and Zhao, Zhe and Gupta, Jai and others},
  booktitle={Advances in Neural Information Processing Systems},
  volume={35},
  year={2022}
}

@article{zhai2024hstu,
  title={Actions speak louder than words: Trillion-parameter sequential transducers for generative recommendations},
  author={Zhai, Jiaqi and Liao, Lucy and Liu, Xing and Vber, Yueming and Li, Boshi and Guan, Celena and others},
  journal={arXiv preprint arXiv:2402.17152},
  year={2024}
}

@inproceedings{geng2022p5,
  title={Recommendation as language processing (RLP): A unified pretrain, personalized prompt \& predict paradigm (P5)},
  author={Geng, Shijie and Liu, Shuchang and Fu, Zuohui and Ge, Yingqiang and Zhang, Yongfeng},
  booktitle={Proceedings of the 16th ACM Conference on Recommender Systems},
  pages={299--315},
  year={2022}
}

@article{zheng2024lcrec,
  title={Adapting large language models by integrating collaborative semantics for recommendation},
  author={Zheng, Bowen and Hou, Yupeng and Lu, Hongyu and Chen, Zhichao and Zhao, Wayne Xin and Wen, Ji-Rong},
  journal={arXiv preprint arXiv:2311.09049},
  year={2024}
}

@inproceedings{kuai2024hourglass,
  title={Breaking the Hourglass Phenomenon of Residual Quantization: Enhancing the Upper Bound of Generative Retrieval},
  author={Kuai, Zhirui and Chen, Zuxu and Wang, Huimu and Li, Mingming and Miao, Dadong and Wang, Binbin and Chen, Xusong and Kuang, Li and Han, Yuxing and Wang, Jiaxing and Tang, Guoyu and Liu, Lin and Wang, Songlin and Zhuo, Jingwei},
  booktitle={Proceedings of the 2024 Conference on Empirical Methods in Natural Language Processing: Industry Track},
  pages={677--685},
  year={2024}
}

@article{chen2026gti,
  title={Grounded Token Initialization for New Vocabulary in LMs for Generative Recommendation},
  author={Chen, Daiwei and Fu, Zhoutong and Jiang, Chengming and Zhang, Haichao and Zhou, Ran and Wang, Tan and Yao, Chunnan and Li, Guoyao and Cai, Rui and Cao, Yihan and Jiang, Ruijie and Borisyuk, Fedor and Shen, Jianqiang and Wu, Jingwei and Vinayak, Ramya Korlakai},
  journal={arXiv preprint arXiv:2604.02324},
  year={2026}
}

@inproceedings{minixhofer2022wechsel,
  title={{WECHSEL}: Effective initialization of subword embeddings for cross-lingual transfer of monolingual language models},
  author={Minixhofer, Benjamin and Paischer, Fabian and Rekabsaz, Navid},
  booktitle={Proceedings of the 2022 Conference of the North American Chapter of the Association for Computational Linguistics},
  year={2022}
}

@article{su2026static,
  title={Vectorizing the Trie: Efficient Constrained Decoding for {LLM}-based Generative Retrieval on Accelerators},
  author={Su, Zhengyang and Katsman, Isay and Wang, Yueqi and He, Ruining and Heldt, Lukasz and Keshavan, Raghunandan and Wang, Shao-Chuan and Yi, Xinyang and Gao, Mingyan and Dalal, Onkar and Hong, Lichan and Chi, Ed and Han, Ningren},
  journal={arXiv preprint arXiv:2602.22647},
  year={2026}
}

@article{hou2025rpg,
  title={Generating Long Semantic {IDs} in Parallel for Recommendation},
  author={Hou, Yupeng and Li, Jiacheng and Shin, Ashley and Jeon, Jinsung and Santhanam, Abhishek and Shao, Wei and Hassani, Kaveh and Yao, Ning and McAuley, Julian},
  journal={arXiv preprint arXiv:2506.05781},
  year={2025}
}

@article{pinclip2026,
  title={PinCLIP: Large-scale Foundational Multimodal Representation at Pinterest},
  author={Beal, Josh and Kim, Eric and Rao, Jinfeng and Wu, Rex and Kislyuk, Dmitry and Rosenberg, Charles},
  booktitle={Proceedings of The ACM Web Conference},
  year={2026}
}

@article{grattafiori2024llama,
  title={The llama 3 herd of models},
  author={Grattafiori, Aaron and Dubey, Abhimanyu and Jauhri, Abhinav and Pandey, Abhinav and Kadian, Abhishek and Al-Dahle, Ahmad and Letman, Aiesha and Mathur, Akhil and Schelten, Alan and Vaughan, Alex and others},
  journal={arXiv preprint arXiv:2407.21783},
  year={2024}
}

@article{kamath2025gemma,
  title={Gemma 3 technical report},
  author={Kamath, Aishwarya and Ferret, Johan and Pathak, Shreya and Vieillard, Nino and Merhej, Ramona and Perrin, Sarah and Matejovicova, Tatiana and Ram{\'e}, Alexandre and Rivi{\`e}re, Morgane and Rouillard, Louis and others},
  journal={arXiv preprint arXiv:2503.19786},
  volume={4},
  year={2025},
  publisher={ArXiv}
}

@article{shao2024deepseekmath,
  title={Deepseekmath: Pushing the limits of mathematical reasoning in open language models},
  author={Shao, Zhihong and Wang, Peiyi and Zhu, Qihao and Xu, Runxin and Song, Junxiao and Bi, Xiao and Zhang, Haowei and Zhang, Mingchuan and Li, YK and Wu, Yang and others},
  journal={arXiv preprint arXiv:2402.03300},
  year={2024}
}

@article{chen2026does,
  title={Does reinforcement learning really incentivize reasoning capacity in llms beyond the base model?},
  author={Chen, Zhiqi and Lu, Rui and Zhao, Andrew and Wang, Zhaokai and Yue, Yang and Song, Shiji and Huang, Gao},
  journal={Advances in Neural Information Processing Systems},
  volume={38},
  pages={57654--57689},
  year={2026}
}

@article{verma2026orbit,
  title={ORBIT: Preserving Foundational Language Capabilities in GenRetrieval via Origin-Regulated Merging},
  author={Verma, Neha and Mehta, Nikhil and Wang, Shao-Chuan and Zhang, Naijing and Tsai, Alicia and Wei, Li and Heldt, Lukasz and Hong, Lichan and Chi, Ed and Yi, Xinyang},
  journal={arXiv preprint arXiv:2605.12419},
  year={2026}
}

@article{kaplan2020scaling,
  title={Scaling laws for neural language models},
  author={Kaplan, Jared and McCandlish, Sam and Henighan, Tom and Brown, Tom B and Chess, Benjamin and Child, Rewon and Gray, Scott and Radford, Alec and Wu, Jeffrey and Amodei, Dario},
  journal={arXiv preprint arXiv:2001.08361},
  year={2020}
}

@String{Computing = "Computing" }

@String{Computer = "{IEEE} Computer" }

@ArtifactSoftware{R,
    title = {R: A Language and Environment for Statistical Computing},
    author = {{R Core Team}},
    organization = {R Foundation for Statistical Computing},
    address = {Vienna, Austria},
    year = {2019},
    url = {https://www.R-project.org/},
}

\end{document}